\documentclass[sigconf,nonacm]{acmart}

\renewcommand\footnotetextcopyrightpermission[1]{}

\usepackage{booktabs}
\usepackage{tabularx}
\usepackage{graphicx}

\newcommand{\systemname}{\textsc{SkillLens}}
\newcommand{\performanceview}{Performance}
\newcommand{\summaryview}{Summary}
\newcommand{\contextview}{Context}
\newcommand{\refinementview}{Refinement}

\title{\systemname{}: An Interactive Human-LLM System for Diagnosing and Refining Agent Skills}

\author{Shuo Yan}
\email{10245102482@stu.ecnu.edu.cn}
\affiliation{%
  \institution{Florida State University}
  \department{Department of Computer Science}
  \city{Tallahassee}
  \state{Florida}
  \country{USA}
}

\author{Rui Sheng}
\email{rshengac@connect.ust.hk}
\affiliation{%
  \institution{The Hong Kong University of Science and Technology}
  \department{Department of Computer Science and Engineering}
  \city{Hong Kong}
  \country{China}
}

\author{Xiaoyan Gu}
\email{xiaoyanGu@zju.edu.cn}
\affiliation{%
  \institution{Florida State University}
  \department{Department of Computer Science}
  \city{Tallahassee}
  \state{Florida}
  \country{USA}
}

\author{Lian Lian}
\email{lianl@wustl.edu}
\affiliation{%
  \institution{Washington University in St. Louis}
  \department{Department of Computer Science \& Engineering}
  \city{St. Louis}
  \state{Missouri}
  \country{USA}
}

\author{Chengran Xu}
\email{xuchengran2004@gmail.com}
\affiliation{%
  \institution{University of Wisconsin--Madison}
  \department{College of Letters \& Science}
  \city{Madison}
  \state{Wisconsin}
  \country{USA}
}

\author{Yifang Wang}
\email{yifang.wang@fsu.edu}
\affiliation{%
  \institution{Florida State University}
  \department{Department of Computer Science}
  \city{Tallahassee}
  \state{Florida}
  \country{USA}
}

\renewcommand{\shortauthors}{Yan et al.}

\begin{document}

\begin{teaserfigure}
  \includegraphics[width=\textwidth]{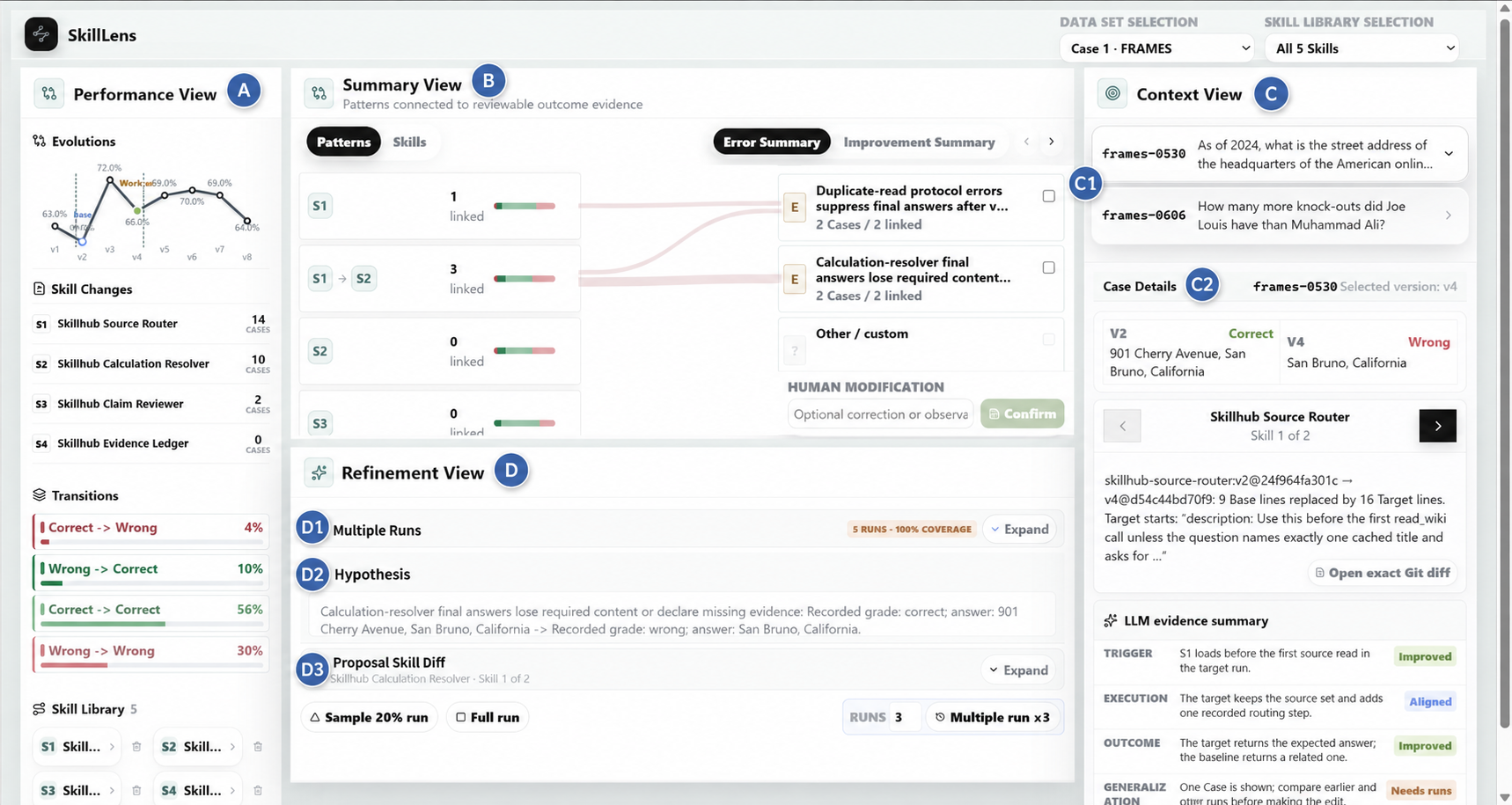}
  \caption{\systemname{} connects performance changes to summarized error patterns, case evidence, and a bounded refinement plan.}
  \Description{A wide screenshot of \systemname{} showing Performance, Summary, Context, and Refinement views connected by skill and case evidence.}
  \label{fig:teaser}
\end{teaserfigure}

\begin{abstract}
LLM agents increasingly rely on reusable skills, yet maintaining them requires developers to connect revisions with changes in behavior and task outcomes. Existing workflows fragment evidence across versions, evaluation cases, and execution traces. We present \systemname{}, an interactive human–LLM system for diagnosing and refining agent skills. A formative study with five developers and researchers identified challenges in prioritizing failures, reconstructing case-level evidence, and preserving context across revisions. \systemname{} links performance changes, case transitions, skill diffs, invocation sequences, traces, and outputs in a recoverable workflow, while keeping model-generated explanations and refinements under human review. 
Through a between-subjects user study with 20 participants, we show that \systemname{} supports developers in diagnosing and refining agent skills through a more usable and less demanding workflow than the baseline. These findings highlight the value of linking version comparisons, case-level evidence, and human-reviewed refinement across skill iterations.

\end{abstract}

\maketitle

\section{Introduction}

LLM agents are increasingly used to carry out recurring, multi-step workflows, where the same procedures, constraints, and tool-use patterns may be needed repeatedly across tasks and sessions. 
As a result, developers are beginning to externalize such procedural knowledge into reusable artifacts rather than specifying it from scratch in every interaction. 
One increasingly common abstraction is the \emph{skill}: a persistent package of instructions, scripts, examples, and supporting resources that an agent can invoke when appropriate. 
Skills are stored in a library available to the agent. When a new request arrives, the agent identifies one or more relevant skills based on the task and the skill's description, loads the corresponding instructions and resources, and follows them while planning actions and producing the final result. 
For example, when asked to turn experimental results into a publication-quality figure, an agent may coordinate a \emph{statistical-review skill} and a \emph{scientific-figure skill}. The former validates the statistical analysis and reporting, while the latter uses these results to design the figure, generate editable plots, and enforce publication requirements~\cite{natureSkillsFigure}. These skills can then be reused across different figure-generation tasks without restating their instructions in every prompt.

Skills themselves are generated in several ways: developers may write them manually, adapt existing skills from public repositories, copy and specialize an internal skill for a new context, or use LLM-based tools to generate and refine them automatically. 
In practice, however, a skill is rarely built successfully in a single attempt or immediately made to work effectively across all relevant tasks. 
For example, a figure-generation skill written for one journal may apply the same formatting rules to a request targeting another venue. The developer then needs to determine whether to revise the shared instructions or create a separate variant for the new requirements. 
Developers therefore have to repeatedly test, revise, and specialize skills as new failures and requirements emerge, often producing multiple related versions and variants over time. 
Existing automated refinement methods~\cite{gautam2026skillaxesharpeningllmauthoredagent,zheng2025skillweaverwebagentsselfimprove} can accelerate this process by optimizing a skill against predefined tasks or evaluation metrics.
However, an improvement on the tasks used for refinement does not by itself establish that the revised skill remains suitable for other tasks or for use alongside other skills. Developers still need to examine regressions and decide whether a revision should be retained~\cite{zhang2026coevolvingskillgenerationpolicy, epperson2025interactivedebuggingsteeringmultiagent}. 
Skill maintenance therefore still requires human-in-the-loop: automated methods can generate alternatives and surface behavioral evidence, while developers must ultimately interpret the trade-offs and decide how an evolving collection of skills should be maintained.

However, involving humans in this process is challenging. 
Developers must reason across many skills, versions, and evaluation cases simultaneously: they need to understand which skill or variant is invoked, how it affects agent execution, which cases improve or regress after an update, and what recurring patterns emerge across iterations. 
These relationships are distributed across skill definitions, execution traces, and evaluation outcomes, making it difficult to organize the evidence needed for diagnosis and maintenance.
To reduce this burden, developers increasingly rely on general-purpose LLMs to summarize traces, compare failures, and identify recurring patterns. 
However, these tools provide little structure for such longitudinal and comparative analysis. 
Their summaries can vary with prompts, supplied context, and conversation history, and developers must repeatedly decide what evidence to provide, what comparisons to request, and how to reconcile observations across iterations. As a result, a more systematic and structured analysis approach is needed.

To address this problem, we conducted a formative study with five participants who had experience developing or maintaining agent skills. 
We examined how they identify failures, inspect execution traces, compare skill versions, use LLMs to summarize cases, and decide what to revise. 
Based on these findings, we present \systemname{}, a system for diagnosing and refining evolving agent skills. \systemname{} organizes the debugging process around the relationship between \emph{skill variants}, \emph{agent behaviors}, and \emph{evaluation outcomes}. It allows developers to compare skill versions across the same evaluation cases, inspect when and how different skills are invoked, trace failures back to relevant skill content and execution steps, and summarize recurring improvements and regressions across cases and iterations.
We evaluate the system in a comparative user study. In summary, this work makes the following contributions:

\begin{itemize}
\item \textbf{An empirical characterization of skill-maintenance workflows and challenges.} Through a formative study with 5 developers and researchers, we characterize how practitioners debug evolving skills, including how they inspect traces, compare versions and cases, use LLM-assisted summaries, and make maintenance decisions. We identify key challenges in maintaining a systematic debugging workflow and reasoning across intertwined skills, versions, and evaluation cases.

\item \textbf{An LLM-assisted interactive system for skill refinement.} We introduce a system, \systemname{}, which seamlessly connects \emph{skill variants}, \emph{agent behaviors}, and \emph{evaluation outcomes} to support systematic diagnosis and refinement across skill iterations.

\item \textbf{A comparative evaluation of skill-refinement workflows.}
Through a between-subjects study with 20 participants, we compare \systemname{} with a baseline system on task completion, completion
time, evidence quality, perceived usability, workload, and task confidence.

\end{itemize}

\section{Related Work}

To position our work, we review research on agent skills and their maintenance. We first discuss how skills have emerged as a reusable abstraction for LLM agents and how researchers have approached skill evolution and management. We then examine human-in-the-loop debugging tools that aim to make agent behavior interpretable and correctable, highlighting their limitations in supporting the longitudinal maintenance of evolving skill libraries.

\subsection{Skills Evolution and Management}

As large language models are increasingly used to carry out multi-step, recurring tasks, researchers have begun to externalize procedural knowledge into reusable units of agent capability. \emph{Skills} are one effective form of this abstraction. A skill is typically defined as a persistent collection of instructions, tool-use patterns, examples, and supporting resources; when facing a new request, the agent retrieves and invokes the relevant skill from a skill library based on the task description \cite{wang2023voyageropenendedembodiedagent, cai2024largelanguagemodelstool}. For example, Voyager accumulates executable code as skills in the open-ended world of Minecraft, enabling the agent to continuously learn new capabilities without human intervention \cite{wang2023voyageropenendedembodiedagent}. LATM, in contrast, treats the language model as a "tool maker" that generates reusable Python tools on demand and later invokes them during reasoning, thereby decoupling tool creation from tool use \cite{cai2024largelanguagemodelstool}. More recent work has extended the notion of skills to multi-agent collaboration and software engineering: MetaGPT and ChatDev assign specialized roles to different agents and follow structured workflows to automatically produce complex software artifacts \cite{hong2024metagptmetaprogrammingmultiagent, qian2024chatdevcommunicativeagentssoftware}. At the same time, as the number of skills grows, an increasing body of research has begun to investigate how skills evolve and how skill libraries can be managed, iterated, and optimized efficiently \cite{yang2026autoskillexperiencedrivenlifelonglearning, zheng2025skillweaverwebagentsselfimprove}.
Recent work has begun to study Agent Skills directly. Large-scale analyses characterize public Skill ecosystems, while SkillsBench and SkillRet evaluate Skill effectiveness and retrieval at scale \cite{ling2026agentskills,li2026skillsbench,cho2026skillret}. Other studies examine when Skills help or fail and how multiple Skills can be selected and ordered for a task \cite{jiang2026demystifying,zhao2026skillcomposer}. These works motivate closer attention to Skill-library quality and use; our focus is on diagnosing and maintaining Skills across versions and runs.

However, most existing research focuses on building or scaling skill libraries from scratch, paying insufficient attention to debugging and maintenance after skills have been created. In practice, the initial version of a skill rarely satisfies all scenarios: its instructions may omit critical assumptions, fail on unseen tasks, or become incompatible with changes in tool interfaces or environmental constraints. As the number of skills increases, overlapping functionality or invocation conflicts among skills can further exacerbate maintenance difficulty \cite{epperson2025interactivedebuggingsteeringmultiagent}. Consequently, developers repeatedly test, revise, and specialize skills, producing multiple versions and variants as new failure cases emerge. Current maintenance approaches fall into two broad categories. The first is manual inspection: developers switch back and forth among raw skill definitions, execution traces, and evaluation results, locating problems by reading logs, running single-point tests, or querying general-purpose language models. This approach is manageable when the skill library is small, but diagnostic efficiency drops sharply as the library scales. The second category comprises automated skill-iteration methods. AutoSkill proposes an experience-driven self-evolution framework that generates new skills from successful trajectories, merges redundant skills, and refines existing ones based on accumulated feedback \cite{yang2026autoskillexperiencedrivenlifelonglearning}. SkillWeaver enables web agents to self-improve by discovering and honing skills \cite{zheng2025skillweaverwebagentsselfimprove}. SkillAxe uses evaluation feedback to guide large language models in automatically polishing generated skills \cite{gautam2026skillaxesharpeningllmauthoredagent}. Other work combines reinforcement learning with skill libraries or jointly optimizes skill generation and policy learning \cite{wang2026reinforcementlearningselfimprovingagent, zhang2026coevolvingskillgenerationpolicy}. While these methods can reduce manual effort to some extent, they tend to be black-box, unstable, require a large number of iterations, and produce outcomes that are difficult to control.
Related work also improves language models and agents through verbal reflection, self-feedback, tool-based critique, and reusable experience \cite{slrw_reflexion2023,madaan2023selfrefine,gou2024critic,zhao2024expel}. In parallel, systems such as DSPy, AFlow, Automated Design of Agentic Systems, and Agent Workflow Memory automate the optimization or construction of language-model pipelines and agent workflows \cite{khattab2024dspy,zhang2025aflow,hu2025adas,wang2025awm}. These approaches automate adaptation or workflow construction; \systemname{} instead supports developers in inspecting how a versioned Skill change affects behavior across Cases and runs.

More importantly, these methods often treat optimization as a one-time objective-function solve, lacking any trace of how skill versions evolve and unable to explain to developers why a particular change causes regressions in some cases. Therefore, human developers must still assume final responsibility: they need to understand the behavioral differences among skill versions in real execution, trade off improvements against regressions, and decide how to maintain an evolving collection of skills \cite{epperson2025interactivedebuggingsteeringmultiagent}.

\subsection{Human--AI Collaboration for Agent Debugging and Maintenance}

Human--AI interaction research provides a complementary perspective on debugging: users need ways to inspect, correct, and guide AI behavior as their understanding of a task develops. Amershi et al.'s guidelines emphasize efficient correction, granular feedback, and communicating the consequences of user actions~\cite{amershi2019guidelines}. For skill maintenance, these principles motivate a division of responsibilities in which an LLM proposes interpretations or revisions, while developers examine the supporting evidence and decide what to change and evaluate.
Related studies also show that explanations must be designed around users' information needs and interpreted with care \cite{liao2020questioning,kaur2020interpreting}. Explanations can increase acceptance of AI recommendations even when they are incorrect, while cognitive-forcing interventions can reduce overreliance \cite{bansal2021whole,bucinca2021trust}. This motivates keeping generated interpretations and revisions open to inspection rather than treating them as conclusions.

Several systems make this collaboration concrete through intermediate inspection and iterative comparison. AI Chains decomposes an LLM task into modular steps that users can modify together with their intermediate results; users can compare alternative strategies and debug unexpected outputs by testing subcomponents~\cite{wu2022aichains}. ChainForge supports comparisons across prompt variations and models for prompt design and hypothesis testing, including iterative refinement~\cite{arawjo2024chainforge}. These systems establish that human--AI collaboration can involve testing and comparing alternatives, rather than merely reviewing a final response. Their emphasis is on prompt chains and response comparisons; our focus is on connecting reusable skill versions to their invocation behavior and case-level outcomes during maintenance.

Agent-focused tools provide complementary forms of inspection and intervention. AutoGen Studio supports configuring and debugging multi-agent workflows~\cite{dibia2024autogenstudionocodedeveloper}, while AgentClick introduces a skill-based human review layer for terminal agents~\cite{zhuang2026agentclickskillbasedhumanintheloopreview}. DiLLS supports interactive diagnosis through hierarchical summaries of agent behavior~\cite{sheng2026dillsinteractivediagnosisllmbased}. Interactive Debugging and Steering and Illuminating LLM Coding Agents further support examining execution behavior through interactive interfaces and visual analytics~\cite{epperson2025interactivedebuggingsteeringmultiagent,wang2025illuminatingllmcodingagents}. Together, these approaches show how developers can inspect execution and intervene at different points in an agent workflow.

Earlier software-debugging research likewise shows the value of linking observed behavior to why/why-not evidence and of supporting the questions developers ask during program changes \cite{ko2004whyline,sillito2008questions}. \systemname{} extends this idea to agent Skills, where relevant evidence is distributed across versions, Cases, invocations, traces, and outputs.

Building on this work, \systemname{} addresses a maintenance task that requires connecting \emph{skill variants}, \emph{agent behaviors}, and \emph{evaluation outcomes}. When a revision improves some cases but regresses others, developers need to compare the affected cases, inspect the corresponding skill changes and invocations, and decide whether a proposed repair warrants another run. We organize these activities into a workflow in which LLM-generated summaries and revision candidates remain open to human confirmation, editing, or rejection. case-level links support inspectable maintenance hypotheses rather than establishing that a skill change caused an outcome. The intended contribution is this connection between version comparison, execution evidence, and human-reviewed revision and validation, rather than interactive debugging or comparative evaluation alone.

\section{Formative Study}
\label{sec:formative-study}

To understand how developers discover, interpret, and correct failures while maintaining reusable agent skills, we conducted a qualitative formative study with five participants (E1-E5) who had experience creating or updating agent skills. Participants were recruited through referrals from expert peers in agent and skill development.
 E1 was a master's student working on agent development; E2 was an undergraduate student in artificial intelligence; E3 was a community contributor and software engineer with reverse-engineering experience; and E4 and E5 were industry engineers. Their experience covered manually authored and patched skills, LLM- or tool-generated initial versions, functional decomposition, composition into larger workflows, and agent-mediated routing. Participants received a US\$20 gift card, following the study protocol. The formative study was covered by the same IRB-approved research protocol, and all participants provided informed consent.

\begin{table}[h]
\centering
\caption{Participant backgrounds and Skill-related experience.}
\label{tab:participants}
\small
\begin{tabular}{@{}p{0.14\columnwidth}p{0.28\columnwidth}p{0.48\columnwidth}@{}}
\hline
\textbf{Participant} & \textbf{Background} & \textbf{Skill-related experience} \\
\hline
E1 & Master's student & Agent development and creating or updating Skills \\
E2 & Undergraduate student in AI & Skill creation, patching, and task-specific debugging \\
E3 & Community contributor and software engineer & Reverse engineering and Skill development \\
E4 & Industry engineer & Creating or updating agent Skills \\
E5 & Industry engineer & Creating or updating agent Skills \\
\hline
\end{tabular}
\end{table}

We used a semi-structured interview guide that asked participants how they recognized problems in skills, how they refined them, and what difficulties they encountered while diagnosing and iteratively revising skills. With participants' consent, the interviews were audio-recorded and transcribed for analysis. 
Then, we conducted an iterative, process-oriented thematic analysis by three researchers. This process produced three connected challenges.
Based on the interviews, we find that the central difficulty of refining skills is not editing a skill in isolation: developers must first determine which errors and skills deserve attention, understand the errors based on specific cases, and determine how to iteratively revise skills across runs.

\subsection{Challenges in Refining Skills}
Through our analysis of the skill refinement process, we identified three interrelated challenges as follows. 

\subsubsection{C1. Connecting Errors to Skills and Prioritizing Inspection}
Developers often need to examine errors together with the skills that may have contributed to them. Although LLM-generated summaries can help suggest potential links between individual errors and problematic skills, these links are often presented in isolation. As a result, developers still lack a clear overview of which skills fail most frequently, which skills are repeatedly associated with particular types of errors, or which combinations of skills tend to fail together. Without such an overview, it is difficult to prioritize which errors to investigate and which skills to inspect first.
Before narrowing down the cause of a failure, developers must first connect the observed error to the relevant skill executions. E5 described relating a skill change to the final result as a time-consuming part of the process, highlighting the effort required to reconstruct these relationships manually. Neither a simple error list nor an aggregate performance score makes such connections explicit, making it difficult for developers to identify recurring problem areas and prioritize their investigation.

\subsubsection{C2. Understanding an Error Through Specific Cases}
Developers often understand an error by tracing what changed in specific cases. They compare the previous and current outcomes with the ground-truth, inspect the corresponding skill difference, and examine how the changed instructions affected routing, intermediate execution, and the final output. 
This process is extremely difficult because it requires developers to integrate evidence scattered across different sources and reconstruct how a skill change propagates through the execution process to produce the observed error (E2 and E5).
E1 stated that, \textit{``Though LLMs can help summarize these records and suggest possible explanations, such explanations must remain grounded in the underlying evidence and need to examine the original cases.''} Developers therefore need an inspectable case-level account that connects skill changes, execution steps, and outcomes.

\subsubsection{C3. Iteratively Refining Skills Across Runs}

Developers refine skills through repeated runs rather than one-off revisions. A change that fixes one case may cause previously correct cases to fail, so developers must compare which errors were resolved, which cases regressed, and whether new problems emerged after each revision. Although LLMs can summarize individual runs or suggest changes, developers still need to determine what to modify next and whether a revision should be kept, revised, or reverted.
This requires developers to reason at two levels. Within the current run, they need to decide which part of the skill should be modified and what change could address the observed failure. At the same time, they must compare the current behavior with the previous run to understand what the last revision actually changed. For example, E3 mentioned that if a routing rule is modified to fix one failed case, they need to check not only whether that case now succeeds, but also whether cases that succeeded previously are now routed differently or begin to fail. Similarly, when a new instruction improves the final output, they still need to determine whether the improvement reflects the intended change or other variation between runs. Skill refinement therefore depends on connecting the current debugging decision with evidence from the previous iteration, rather than evaluating each run in isolation.

\subsection{Design Considerations}

Based on the above three challenges, we derived five design considerations as follows.

\textit{\textbf{DR1. Provide an Overview of Historical Performance and Run Context.}}
A system should help developers understand how skill performance has changed over time and under what conditions each run was produced. It should summarize historical performance together with key run information, such as the evaluated cases, model and tool settings, skill versions, and other relevant metadata. This overview provides the context needed to interpret later improvements and failures.

\textit{\textbf{DR2. Connect New Improvements and Errors to Relevant Skills.}}
A system should help developers identify what improved or failed in a new run and relate these changes to the skills involved. It should surface newly resolved cases, newly introduced errors, and persistent failures, and show which individual skills or skill combinations are associated with them. This helps developers quickly identify where important changes occurred and which skills deserve further inspection.

\textit{\textbf{DR3. Support Understanding Through Original Cases.}}
A system should allow developers to examine the original cases behind summarized improvements, errors, and skill-related patterns. For a selected case, it should bring together the relevant skill differences, execution process, intermediate results, and final outputs. By returning to the underlying cases, developers can better understand how an observed pattern occurred and judge whether the summarized explanation is supported by the actual execution.

\textit{\textbf{DR4. Support Hypothesis Generation, Testing, and Application.}}
A system should help developers turn observed problems into testable hypotheses about how a skill should be changed. Developers should be able to generate and refine candidate explanations or modifications, test them on relevant cases, inspect the resulting behavior, and decide whether the change should be applied. The system should keep the hypothesis, proposed modification, and test results connected throughout this process.

\textit{\textbf{DR5. Support Iterative Exploration and Refinement.}}
A system should support repeated cycles of inspection, modification, and evaluation. After each change, developers should be able to compare the new run with previous runs, examine what improved or regressed, and use this evidence to decide what to explore or modify next. This enables developers to progressively refine skills rather than treating each run as an isolated debugging attempt.

\subsection{Design Exploration: Version-Centric Temporal Canvas}
\label{sec:alternative-design}

\begin{figure}[t]
  \centering
  \includegraphics[width=\columnwidth]{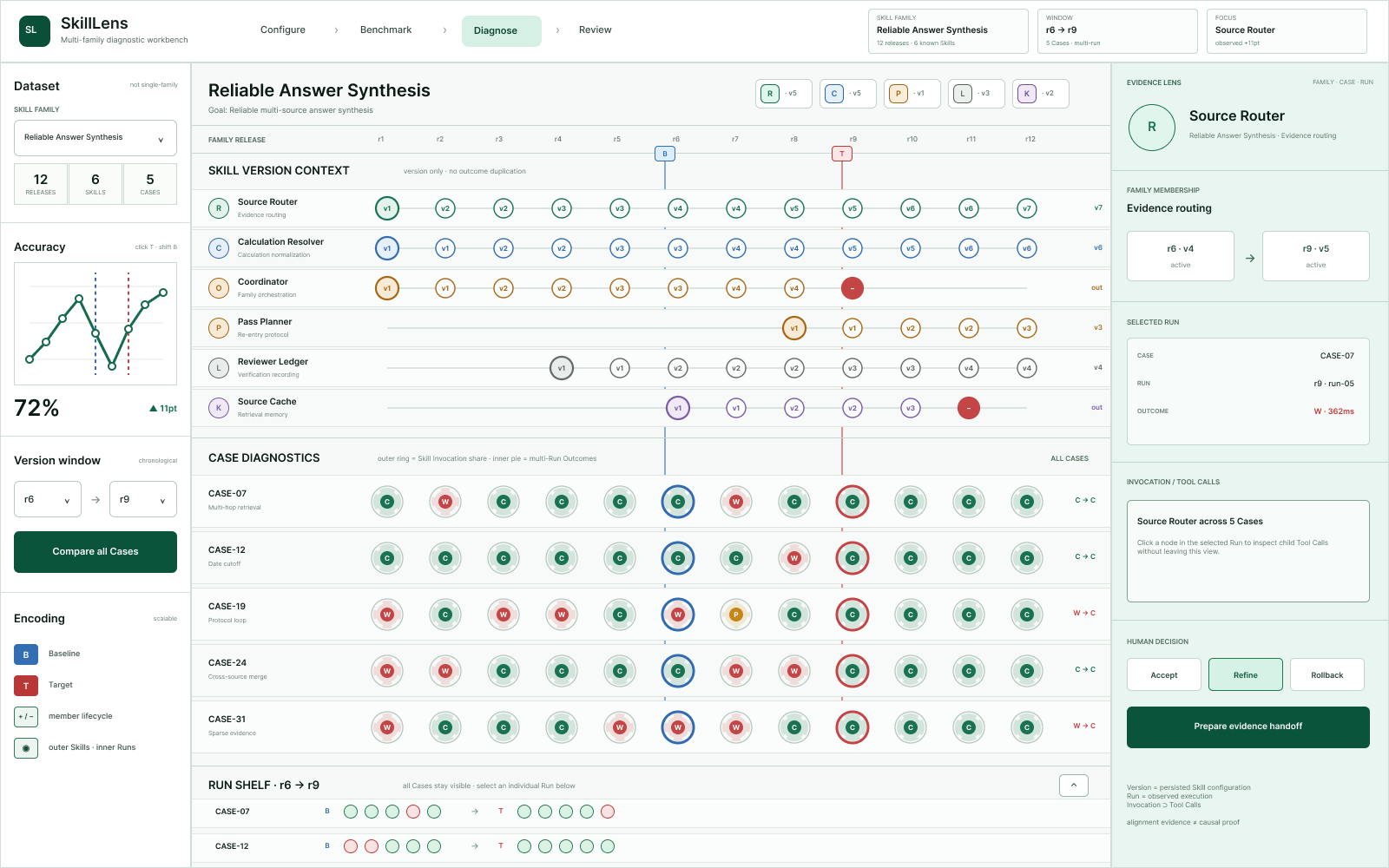}
  \Description{An alternative interface design centered on a skill-version timeline,with case outcomes aligned to versions and a persistent evidence panel.}
  \caption{Explored version-centric temporal canvas. A skill-version timeline is coordinated with case outcomes and a persistent Evidence Lens.}
  \label{fig:alternative-temporal-canvas}
\end{figure}

During early design exploration, we considered a version-centric temporal canvas that placed the skill-version timeline at the center of the workspace, with case outcomes aligned to version changes and a persistent Evidence Lens alongside the timeline. This arrangement made skill evolution and cross-version outcome changes immediately visible, allowing developers to inspect temporal patterns and related evidence within one coordinated surface.
Figure~\ref{fig:alternative-temporal-canvas} shows this explored version-centric temporal canvas.

The main limitation was information density. The timeline, case outcomes, and evidence details competed for attention, while the transition from a detected change to a narrowed diagnostic scope and a maintenance action remained implicit. In particular, the relationship between evidence narrowing and the later hypothesis and what-if workflow was less explicit than required by DR3--DR5. We therefore retained this canvas as a design reference, but converged on the coordinated multiview workspace described in the following section. This decision reflects researcher-led design reasoning and prototype review, rather than a measured user preference.

\section{\systemname{}}
\label{sec:system}

\systemname{} is an interactive diagnostic and refinement workbench for developers of evolving AI-agent skill families. It supports the analysis of outcome changes across skill versions and runs on a shared set of benchmark cases, connecting these changes with skill invocations, execution traces, skill revisions, and output evidence. \systemname{} neither determines which skill family version is the \textit{best} nor automatically attributes an outcome change to a particular revision. Instead, it progressively helps narrow the analysis scope so that developers can formulate refinement hypotheses that are inspectable, revisable, and testable through subsequent runs.

\begin{figure*}[t]
  \centering
  \includegraphics[width=\textwidth]{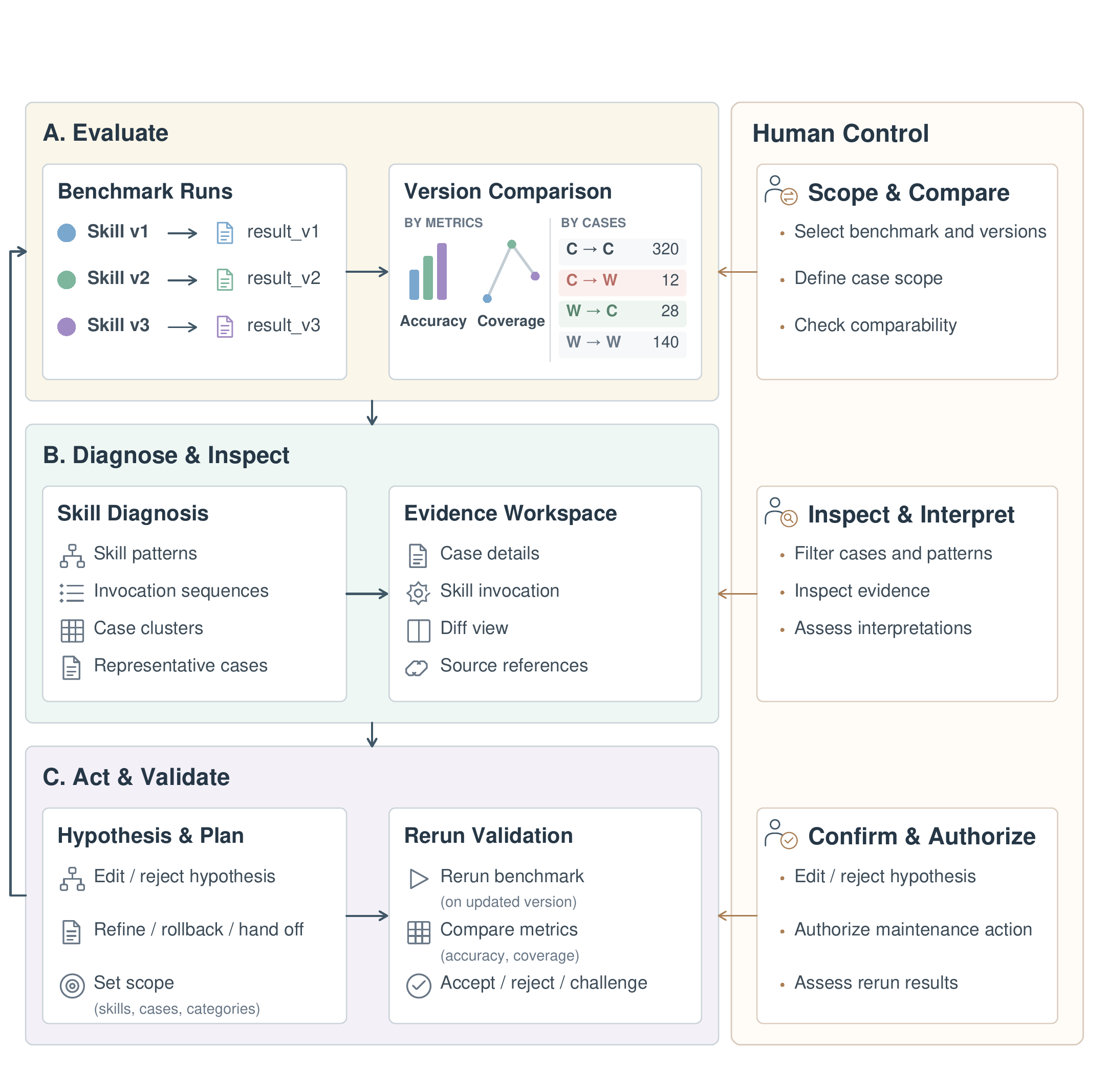}
  \caption{The \systemname{} refinement workflow: comparing
    benchmark results, inspecting case and skill evidence,
    and validating human-authorized revisions through reruns.}
  \Description{Three connected stages: Evaluate, Diagnose
    and Inspect, and Act and Validate. Human review guides
    each stage, and rerun results feed the next iteration.}
  \label{fig:system-pipeline}
\end{figure*}

Figure~\ref{fig:system-pipeline} summarizes the maintenance
workflow of \systemname{}. Developers compare benchmark
results across different skill family versions, inspect affected cases and execution evidence, and formulate revisions for subsequent
validation. Human review guides scope selection, results interpretation, and authorization throughout the process, while rerun results inform the next iteration.

We implement this workflow in our system, \systemname{} (Fig.~\ref{fig:teaser}).
% Figure~\ref{fig:teaser} shows our system overview that support this workflow. 
% support this workflow, with dataset and skill-library
% selectors defining the collection under inspection. 
The \performanceview{} View (Fig.~\ref{fig:teaser}A) presents an overall performance summary across skill iterations. The \summaryview{} View (Fig.~\ref{fig:teaser}B) connects skill-invocation patterns with candidate outcome explanations and supports human correction and confirmation. The \contextview{} View (Fig.~\ref{fig:teaser}C) presents linked cases, version-specific outcomes, skill changes, and an evidence summary. The \refinementview{} View (Fig.~\ref{fig:teaser}D) brings together relevant runs, a hypothesis, and a proposed skill revision for subsequent evaluation.

The four views are linked through a shared comparison context. Selecting runs or versions in the Performance View determines the comparison shown in the Summary View. Selecting a Pattern, Skill, or outcome category in the Summary View narrows the cases shown in the Context View. Developers can return to an earlier view to revise these selections without discarding saved classifications or decisions. 
We introduce the system workflow and each view using a usage scenario, where a developer observes a benchmark regression between two adjacent skill-family versions, locates the affected cases and skills, inspects the corresponding evidence, and formulates a refinement for subsequent validation.

\subsection{\performanceview{} View}
%: Establishing the Comparison Scope
\label{sec:system-performance}

The \performanceview{} View (Fig.~\ref{fig:teaser}A) establishes the benchmark, skill library, and run comparison before diagnosis (\textbf{DR1}). The Evolution chart shows the observed outcome percentage across ordered skill-family versions. When the selected runs contain explicit correctness labels, this percentage corresponds to benchmark Accuracy computed from cases with explicit correctness labels. The number of scored cases and scoring coverage are reported alongside the chart, while run provenance records the benchmark, evaluator, and execution conditions.
The Skill Changes panel lists the skills associated with the selected comparison and reports their affected case counts. The Transitions panel summarizes case-level changes, including correct-to-wrong, wrong-to-correct, correct-to-correct, wrong-to-wrong, and unscored cases. The Skill Library panel lists the available skills and their selected versions.
Developers assign three roles through the chart or version selectors: \textit{Base} denotes the comparison origin, \textit{Target} denotes the run whose outcomes are being inspected, and \textit{Work-on} identifies the version to which a subsequent refinement will be applied. Separating Target from Work-on prevents a developer from unintentionally applying a revision to an older or unrelated skill version while inspecting a historical comparison.

After the version range is selected, users can examine whether the runs use compatible benchmark, case-scope, and evaluation settings. When some cases remain unscored, the view reports both scored and unscored counts and computes benchmark Accuracy only from cases with explicit correctness labels. If the runs cannot support a direct comparison, \systemname{} exposes the incompatibility instead of presenting potentially misleading performance deltas or case transitions. This design makes the scope and evidential basis of a comparison visible before the developer begins interpreting behavioral differences.

Developers can further filter the comparison by outcome transition, such as regressions from correct to incorrect, improvements from incorrect to correct, persistent failures, or unscored cases. In the usage scenario, the developer selects the benchmark dataset, skill library, and runs to compare. The developer assigns the Base run as the comparison reference, the Target run as the run whose outcomes will be inspected, and the Work-on version as the version to which a later refinement may be applied. Selecting a transition or skill carries its case scope to the \summaryview{} View and \contextview{} View.

\subsection{\summaryview{} View}
%: Locating Patterns, Skills, and Cases
\label{sec:system-summary}

%The \summaryview{} View (Figure~\ref{fig:teaser}(B)) organizes cross-version outcome changes into diagnostic entry points (\textbf{DR2}). 
% 1. two modes
The \summaryview{} View (Fig.~\ref{fig:teaser}B) connects case groupings with candidate error and improvement explanations (\textbf{DR2}). The left panel groups cases in two ways. Pattern grouping organizes cases according to observed skill-invocation sequences, such as an S1 $\rightarrow$ S2 sequence. skill grouping organizes cases by the selected skill and reports its associated case counts, transition distribution, and observed exposure across the compared runs. Each grouping retains its case identifiers, transition distribution, and evidence scope.
%Developers can switch between \textit{Error Summary} mode and \textit{Improvement Summary} mode. 
%\textit{Error Summary} mode shows regressions and persistent failures, answering questions such as which problems were introduced or remained unresolved in the newer version. 
%\textit{Improvement Summary} mode focuses on cases that became correct, allowing developers to inspect where a revision may be associated with beneficial outcomes. 
The two modes retain the same visual design and interaction structure, enabling developers to explore current problems as well as potential gains and risks beyond a single aggregate accuracy value.

% 2. visual encoding: what is on the left div, what is on the right div, and what does the flow mean? color and thickness? 
Developers often need to inspect errors and skills together, but a free-form LLM summary can leave their relationships fragmented, making it unclear which skill is associated with which error and which error to inspect first. 
The \summaryview{} View externalizes these links and keeps the linked cases available, rather than requiring the developer to reconstruct them from text on every turn.
% 2-1 what is on the left div
The left panel provides two case-grouping methods. Pattern grouping is derived from contiguous ordered skill-invocation sequences recorded during execution. For example, cases in an S1 $\rightarrow$ S2 group share the observed order of these skill calls. Skill grouping instead collects cases associated with an individual skill and reports its exposure and transition distribution across the compared runs. These groupings summarize observed execution patterns; they are not semantic or causal explanations.
%The left part shows ... 
%\systemname{} provide two analysis methods to dive into the skill-calling patterns: 
% \systemname{} provides two complementary analysis modes. 
%\textit{Pattern} method organizes cases according to their outcome transitions and observed skill-invocation characteristics, helping developers identify cases with similar execution phenomena. 
%For example, one group of regressed cases may share a missing skill invocation, whereas another may invoke the same skill but encounter a later tool-execution failure. 
%These patterns are candidate evidence groupings rather than automatically %inferred causes. 
%\textit{Skill} mode instead summarizes individual skills and presents their individual-level invocation coverage, version changes, and associated numbers of improved or regressed cases.
% 
% 2-2 what is on the right div
The right panel first presents an \textit{Error Summary}. In this mode, it lists candidate explanations for regressed or persistently failing cases, generated from the selected case evidence, including skill changes, invocations, traces, outputs, and evaluation records. The flow connects those explanations to the corresponding Pattern or skill groups. A category may describe, for example, missing content in a final answer or a failure that appears after a particular skill invocation. These entries are provisional classifications for human review. They do not establish that a skill caused the observed outcome.

The flow in the middle connects each left-side case grouping to the candidate categories assigned to its cases. Its thickness represents the number of cases in the displayed connection. The source and destination positions identify the connected grouping and candidate category. The flow summarizes an evidence association between a case grouping and a candidate explanation. It does not represent execution order or causal influence.
We also provide an \textit{Improvement Summary} for cases that changed from wrong to correct or otherwise showed a favorable outcome transition. In this mode, the left panel uses the same Pattern and skill groupings, while the right panel lists candidate improvement explanations generated from the corresponding case evidence. The flow represents the cases associated with each grouping and improvement category. Developers can therefore inspect potential gains together with the evidence supporting those cases.
%For the error analysis, we use a Sankey-style diagram to show how cases connect visible skill-invocation patterns with candidate error categories in the summary (Figure~\ref{fig:teaser}(B)). The source and destination positions identify the connected pattern and category. Ribbon width encodes the number of cases assigned to that connection in the displayed primary flow. These links represent case associations, not execution order or causal relationships. Selecting a pattern highlights its associated categories, and selecting a category narrows the cases available in the \contextview{} View.

In the usage scenario, the developer clicks a Pattern group on the left panel, such as S1 $\rightarrow$ S2. The flow is highlighted to show the relevant cases from that invocation group and the corresponding error types on the right. The developer then selects an error type to narrow the case scope passed to the Context View. After reviewing the Error Summary, the developer switches to Improvement Summary to inspect cases that improved and compare their associated evidence.

%In the running example, the developer observes that many regressed cases are concentrated in one pattern associated with two modified skills. The developer selects a representative case and proceeds to the 
%\contextview{} View to determine whether the observed association is supported by the underlying execution records.

\subsection{\contextview{} View}
%: Inspecting Case-Level Behavior
\label{sec:system-context}

The \contextview{} View presents the cases passed from the Performance or Summary View and supports case-level evidence reconstruction(DR3). The case List shows the selected transition, Pattern or skill grouping, and case scope. The Case Details area presents the selected case and places the Base and Target evaluation states and outputs side by side. This allows the developer to first establish what changed before examining related execution behavior. An unscored case is explicitly shown as \textit{Unscored}; successful execution or the presence of an output is not treated as evidence of correctness.

Below the outcome comparison, \systemname{} presents the skill calls recorded in the two runs. The system distinguishes \textit{trigger evidence} from \textit{execution evidence}. Trigger evidence indicates whether a skill entered the execution context or was selected through routing, whereas execution evidence indicates whether it was invoked, its order, and its subsequent tool calls. This distinction helps developers separate a skill that was not triggered from one that was triggered but did not execute effectively. Invocation sequences are ordered according to the recorded events, enabling comparison of added, missing, reordered, or repeated calls.

Developers can expand three complementary evidence sources. \textit{Skill Diff} presents textual changes between the Base and Target versions and provides access to the complete skill snapshots, preventing a developer from interpreting a revision solely from a truncated summary. \textit{Trace} presents intermediate retrieval, calculation, tool-use, and answer-generation events. \textit{Output Evidence} places final outputs and their evaluation outcomes side by side. Aligning these sources around the same case allows the developer to examine whether an aggregate pattern is consistent with the recorded invocations and outputs.

\systemname{} retains provenance for each evidence item. Invocations and traces captured by the execution environment are marked as recorded; displays reconstructed from invocation events are marked as derived; LLM-generated explanations are explicitly presented as suggestions; and logs, traces, or skill snapshots that were not retained are marked as unavailable. Developers may request an LLM-generated summary of the current case, but this summary neither replaces the underlying records nor automatically becomes a maintenance hypothesis. Progressive disclosure keeps outcome and invocation changes visible first, while complete logs remain available on demand.

In the usage scenario, the developer observes that a skill invocation present in the Base run is absent from the Target run and that the corresponding skill text was also revised. Together, these observations form a candidate explanation worth investigating, but \systemname{} does not claim that the textual revision caused the regression. The developer then adds the case and related skill to a human classification.

\subsection{\refinementview{} View}
\begin{figure*}[t]
  \centering
  \includegraphics[width=\textwidth]
    {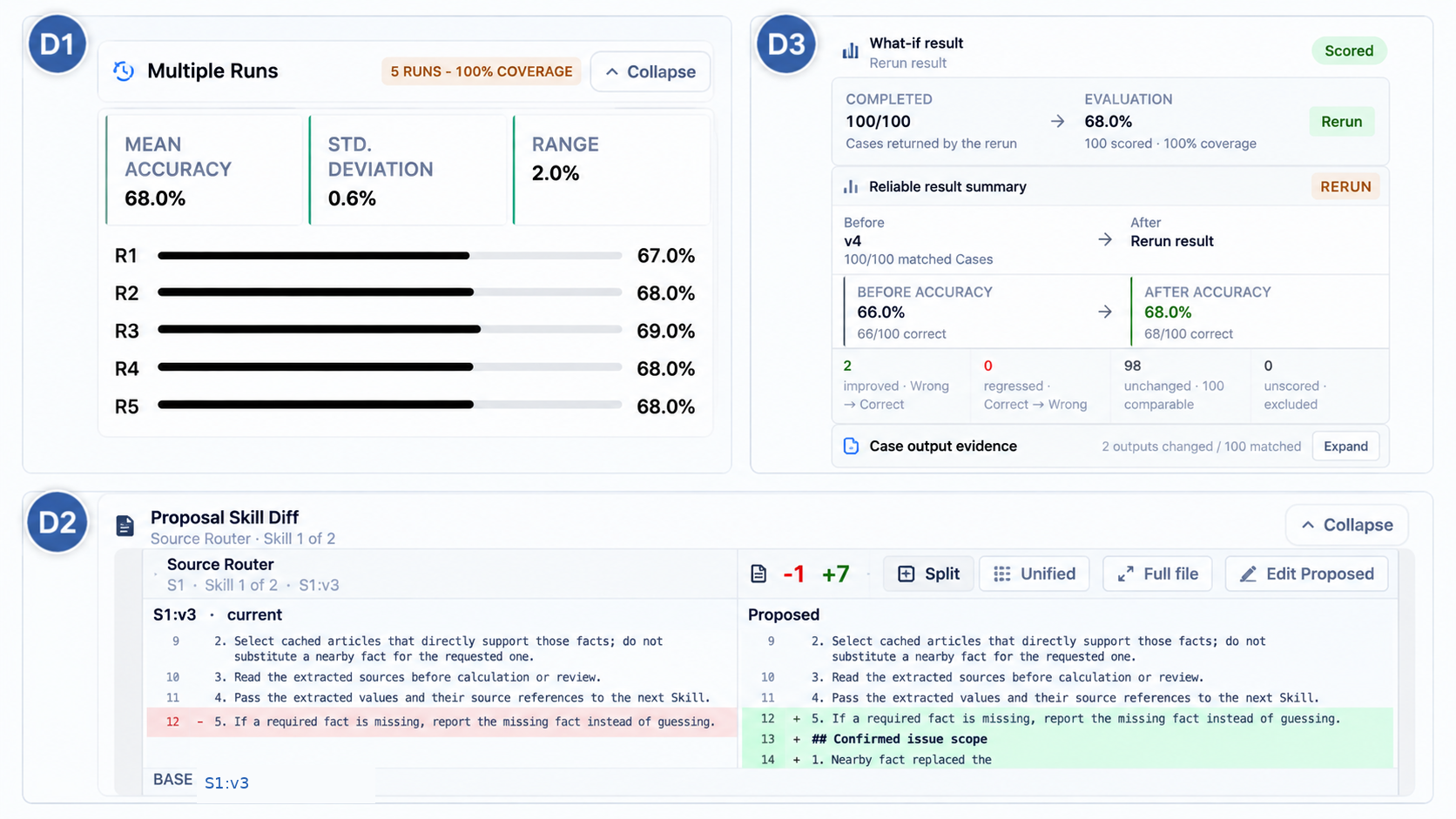}
  \caption{Details of the Refinement View:
    (D1) Multiple Runs provide matched historical and repeated-run context;
    (D2) Proposal Skill Diff supports inspection and editing of the proposed Skill revision;
    (D3) What-if Result reports the rerun outcome and before-and-after case transitions. }
  \Description{A composite interface figure showing
  version history, a proposed skill diff, and a rerun
  result with before-and-after accuracy and case transitions.}
  \label{fig:refinement-details}
\end{figure*}

%: From Evidence to Maintenance Action
\label{sec:system-refinement}
Figure~\ref{fig:refinement-details} details the Multiple Runs, Proposal Skill Diff, and What-if Result regions of the Refinement View.

The \refinementview{} View (Fig.~\ref{fig:teaser}D) transforms evidence inspection into a human-controlled maintenance process (\textbf{DR4}). Developers first select the cases relevant to the current problem and review suggested failure categories, such as missing triggering, problematic skill content, invalid tool arguments, or model noncompliance. They may confirm, rename, merge, split, or reject these categories, while cases with insufficient evidence can remain unclassified. \systemname{} stores the human-authored classification together with its case and evidence references, allowing subsequent revisions to be traced back to the judgment on which they were based.

%After classification, the Hypothesis region (Figure~\ref{fig:teaser}(D2)) records the human-reviewed explanation and scope. \systemname{} can then generate a candidate refinement from the selected cases, skills, and problem categories, while the Proposal Skill Diff region (D3) exposes the proposed textual change. The candidate is presented as structured Markdown containing the confirmed issue scope, original skill content, and proposed revision. developers can directly edit the proposal, request another suggestion, or specify that a skill should remain unchanged. The LLM is limited to generating and revising candidate text; it cannot confirm a failure attribution, overwrite a skill file, or initiate a validation run without human authorization.
After classification, the Hypothesis region in the Refinement View records the human-reviewed explanation and scope. The Proposal Skill Diff region presents the current and proposed Skill versions side by side for inspection and editing. Developers can revise or reject the proposal before choosing a bounded validation plan.

%After confirming the refinement hypothesis, developers use the rerun controls in Figure~\ref{fig:teaser}(D3) to choose among three What-if scopes.
After confirming the refinement hypothesis, developers choose among three What-if scopes: a Sample Run, a Full Run, or Multiple Runs. The Multiple Runs region (D1) keeps matched historical and repeated executions available for comparison, while the What-if result panel reports the outcome of the selected rerun.
A \textit{Sample Run} provides a lower-cost initial check on a subset of cases. A \textit{Full Run} evaluates the complete benchmark scope. \textit{Multiple Runs} repeat the execution under matched model, case-scope, evaluator, and environment conditions. The Multiple Runs region (Fig.~\ref{fig:teaser}D1) keeps the matched historical and repeated executions available for comparison. Multiple Runs help determine whether an improvement or regression persists across repeated executions or may reflect run-to-run sampling variation. A sequence of historical skill versions is therefore not treated as a set of statistical replicates.

When an execution completes, \systemname{} presents the before-and-after evaluation outcomes, scoring coverage, and case transitions and allows developers to reopen the evidence for individual cases. For Multiple Runs, each replicate is retained separately rather than being replaced by a single aggregate value. Only condition-matched cases with explicit evaluation labels enter the comparison statistics, while unscored outcomes remain visible through coverage reporting. Based on this evidence, developers choose \textit{Accept}, \textit{Reject}, or \textit{Challenge}. Accept indicates that the result can inform the next version; Reject discards the current refinement; and Challenge records that the evidence remains insufficient and that the hypothesis or validation scope should be revised.

After each run, developers compare which previously failing cases improved, which previously correct cases regressed, and whether new errors appeared. They repeat the same scope check, evidence review, hypothesis revision, bounded run, and cross-run comparison workflow so that each iteration is comparable rather than ad hoc (\textbf{DR5}).

For example, the illustrated summary links a skill sequence to a candidate category concerning premature declarations of insufficient evidence. Developers can inspect a linked case in the \contextview{} View and return to the Human Modification field to enter a correction or observation. Selecting Confirm records the reviewed interpretation together with the selected cases and skills for subsequent refinement. Merely selecting a category does not confirm its explanation.

\subsection{Implementation}
\label{sec:system-implementation}

The \systemname{} frontend is implemented with React and TypeScript, with a FastAPI backend. Stable identifiers for runs, comparisons, cases, skills, and evidence connect the views and preserve a shared comparison context during filtering and drill-down. The backend imports benchmark outcomes, skill snapshots, invocation events, traces, and outputs while retaining their provenance. Human classifications, hypothesis decisions, and version feedback are stored as append-only records to preserve the history of the maintenance process.

LLMs are used to generate editable case explanations and refinement suggestions. After human confirmation, a What-if task can be executed in an isolated Codex environment and imported into \systemname{} as a separate run. A result is attached to the originating hypothesis only when its version, case scope, and evaluation settings satisfy the comparison requirements. This implementation supports the complete interaction workflow from benchmark changes to evidence inspection and validation while preserving human control and evidence provenance.

\section{Evaluation}
\label{sec:evaluation}

We conducted a between-subjects controlled user study to examine whether
\systemname{} supports the refinement needs identified in our formative
study: connecting errors with relevant skills, reconstructing failures
through case-level evidence, and modifying skills and assessing their
effects across multiple execution rounds. We also examined the usability
and workload of this refinement workflow. The study compared
\systemname{} with a baseline toolchain using the same initial materials.
It evaluated the complete workflows through which developers inspected
evidence, implemented revisions, and assessed subsequent outcomes.

% Our design draws on the time-bounded diagnostic task used in DiLLS~\cite{slrw_dills2026}, while extending the task from diagnosing recorded failures to revising versioned skills and examining the resulting runs.

We organized the evaluation around three research questions corresponding to the formative study's maintenance needs:
\begin{description}
    \item[RQ1:] Does \systemname{} help developers identify skills or skill sequences associated with observed errors and prioritize further inspection?
    \item[RQ2:] Does \systemname{} help developers reconstruct an error through case-level evidence and form a supported, appropriately scoped explanation?
    \item[RQ3:] Does \systemname{} help developers implement skill revisions, assess their effects across multiple execution rounds, and use the resulting evidence to decide to retain, further revise, or roll back the changes?
\end{description}
Task quality and completion are the primary outcomes. Completion time and subjective measures characterize the effort associated with the workflow. For RQ3, task records and interviews are also used to examine how participants interpreted the results of later runs and how those results affected their final decision.

\subsection{Participants}

We recruited participants with experience in AI agents, skills, benchmarks, execution-log analysis, or related debugging tasks. The study included 20 participants, with 10 assigned to each condition. Participants were stratified by gender and background, and randomly assigned to \systemname{} or Baseline within each stratum. Background strata consisted of junior and senior participants. Each participant took part in one formal session, which included the required execution rounds. Each condition comprised eight male and two female participants. The study was conducted with IRB approval. All participants provided informed consent and received a US\$20 Amazon gift card as compensation.

\subsection{Study Materials and Conditions}

We used FRAMES to instantiate a skill-refinement task requiring comparison across versions and inspection of individual cases. FRAMES (Factuality, Retrieval, And reasoning MEasurement Set) evaluates retrieval-augmented systems on factuality, retrieval, and reasoning~\cite{krishna2024factfetchreasonunified}. The dataset contains 824 multi-hop questions requiring information from 2--15 Wikipedia articles. We used a fixed subset of 100 benchmark cases and the v1--v8 version history of one web-search skill family. These tasks require repeated retrieval, source reading, information integration, and answer verification, providing opportunities to inspect how skill revisions relate to execution behavior and outcomes.

The shared study scenario, \textit{Case 1--FRAMES}, contained the historical skill versions, benchmark results, case transitions, skill invocations and sequences, traces, outputs, and source records. Case 1 refers to the complete scenario, which includes multiple benchmark cases. All participants began with access to the same historical versions and evidence. They selected a Base--Target interval and a change they considered worth investigating. We assessed whether their selection and subsequent actions were supported by the materials, rather than requiring a predetermined interval, skill, or maintenance direction. New skill versions and runs produced during the task were distinguished from the historical v1--v8 records.

We constructed two conditions to compare how the workflows supported the same refinement task. In the \systemname{} condition, participants used Performance to inspect version-level results and case transitions, Summary to narrow the investigation to skills, sequences, and cases, and Context to examine skill diffs, traces, outputs, and sources. They used Refinement to review hypotheses and candidate revisions, edit skill content, initiate What-if executions, and inspect the resulting evidence before making a maintenance decision. In the Baseline condition, participants inspected the initial materials through flat tables, raw records, and a preconfigured chat-based toolchain. They modified skills, initiated evaluations through the provided baseline workflow, and recorded their reasoning and decisions in the common response form. The Baseline did not provide \systemname{}'s linked views or integrated maintenance states.

Both groups received the same task instructions, formal-task time limit, initial benchmark materials, and researcher support. Runs created during the task were recorded with their associated skill version and case scope. The comparison concerns the complete workflows, rather than isolating the contribution of an individual visual or generative component. Participants' revisions could produce different subsequent outputs; equal access to the initial evidence did not require identical evidence after execution.

\subsection{Task and Procedure}

The formal task required participants to investigate a selected change, implement a skill revision, complete multiple execution rounds, and submit an evidence-supported refinement decision.
They could inspect the materials in any order, but the submission had to cover a prioritized change (version interval, related skill or skill sequence, representative case, and reason for prioritizing it), evidence that supported, qualified, or contradicted the observation (with each item's location, content, and role, plus an alternative or potentially falsifying explanation), and a maintenance decision (reviewed hypothesis, implemented revision and diff, validation scope, run outcomes, and a justified choice to retain, further revise, or roll back the change).
A What-if plan specified the intended test before execution; submitting the plan alone did not complete the task.

Each session began with an approximately 15-minute condition-specific tutorial and practice task using materials separate from the formal scenario. Both groups learned how to record evidence locations and distinguish an observed association from an established cause. A comprehension check assessed their understanding of the task and evidence boundary and did not contribute to the formal outcomes. The researcher then opened the formal materials and common response form and started the timer. The formal task had a 40-minute limit. Participants worked independently and could think aloud. The researcher could clarify instructions or resolve technical problems but did not suggest an interval, skill, case, or maintenance direction. At the deadline, the current submission and execution status were retained regardless of completion.

Skill editing and multiple execution rounds were required parts of the formal task. Participants inspected subsequent outcomes, compared improvements and regressions within the evaluated scope, and used this evidence to justify their final decision. Each run was associated with the skill version and case scope used. Repeated executions of an unchanged version were distinguished from executions following a further revision: the former examine run-to-run variation, whereas the latter document changes across revisions. Historical runs did not count toward the required task executions.

After the task, participants completed the workload and usability questionnaires and took part in a semi-structured interview organized around the three research questions. The interview examined why they selected a change, which evidence supported or weakened their explanation, how they chose a revision, and how the execution results affected their final decision. It also elicited difficulties in inspecting evidence, modifying skills, and interpreting subsequent runs. 

\subsection{Metrics and Analysis Methods}
We report task completion, evidence quality, and completion time as separate outcomes. Task completion records whether a participant completed the required refinement workflow within the task window. Evidence quality characterizes the traceability, relevance, and scope of the submitted reasoning, but does not determine whether the task was completed. RQ1 is examined through selected changes, related skills, representative cases, and prioritization reasons; RQ2 through submitted evidence and explanations; and RQ3 through the revision made by each participant, the results of later runs, and how those results were used to justify the final decision.

\textit{Evidence quality.} Two coders independently score evidence locations and explanations on a 0--2 rubric grounded in the study materials: 0 if the evidence is untraceable or unrelated to the selected change; 1 if at least one relevant item is present but the qualifying evidence or explanation is incomplete or inaccurate; and 2 if supporting and qualifying evidence are traceable, their roles are explained accurately, and the judgment is appropriately bounded. Alternative explanations or proposed falsification tests are recorded separately from observed counterevidence. Scores depend on the evidence and reasoning, not on interface fields or text length.

\textit{Task completion.} A submission was classified as complete when the participant completed the required workflow components within the task window: selecting a change, implementing a skill revision, completing the required execution rounds, and recording a final maintenance decision. Evidence-quality scores were analyzed as a separate outcome and were not used to determine task completion. A completed submission could therefore receive any evidence-quality score, whereas a plan-only submission or an unfinished execution was classified as incomplete.

\textit{Revision and execution outcomes.} For RQ3, we examine what participants changed, what happened in later runs, and how they used those results when making the final decision. We consider whether participants checked both improvements and regressions before retaining a revision, and whether the decision was supported by evidence beyond the case or cases that motivated the change. We also examine whether a favorable rerun was treated as sufficient confirmation of the earlier explanation when other evidence remained unresolved. These observations are summarized descriptively rather than combined into a single score. Unchanged-version repetitions are kept separate from successive revisions. Benchmark Accuracy uses only cases with explicit correctness labels and is reported with scoring coverage; unscored cases are excluded from the denominator.

\textit{Time.} Completion time is the interval from task start to submission for completed tasks. Incomplete tasks have no recorded completion time and are excluded from time summaries.

\textit{Subjective measures.} After the task, participants completed the System Usability Scale (SUS), an adapted seven-point NASA-TLX questionnaire, and four task-confidence items. SUS totals use the standard 0--100 scoring formula and are interpreted against the commonly used normative reference mean of 68. NASA-TLX ratings were analyzed separately for Mental Demand, Physical Demand, Temporal Demand, Performance, Effort, and Frustration. Each dimension was rated on a seven-point scale. The Performance item had already been direction-adjusted during data collection, so no additional reverse scoring was applied. We did not combine the six dimensions into an overall workload score.
The confidence items cover performance-change identification, case-level evidence localization, evidence-supported hypothesis formation, and maintenance planning.

\textit{Analysis.} Task completion, evidence quality, completion time, and task-confidence ratings were summarized descriptively. NASA-TLX dimension scores were compared between the \systemname{} and Baseline conditions using Welch's independent-samples t-tests. Because six NASA-TLX dimensions were tested, the resulting p-values were adjusted using Holm's method. The performance item was analyzed in its original direction, and no composite workload score was calculated. SUS total scores were compared between the \systemname{} and Baseline conditions using Welch's independent-samples t-test. For each inferential comparison, we report condition means and standard deviations, the mean difference with a 95\% confidence interval, Welch's $t$ statistic, degrees of freedom, p-value, and effect size.

\section{Results}
\label{sec:results}

We report task completion, completion time, evidence quality,
and subjective experience for 20 participants, with 10 in the
\systemname{} condition and 10 in the Baseline condition.
We also examine how participants used the results of later runs
when making their final maintenance decisions. Task performance,
evidence quality, and these decision records are summarized
descriptively.

\subsection{Task Completion and Time}

Nine of ten participants completed the task in each condition.
The remaining participant in each condition was marked as incomplete and had no recorded completion time.
Task completion was determined independently of the evidence-quality score.

Among participants who completed the task, the median
completion time was 26 minutes with \systemname{} and 32 minutes
with Baseline. Mean completion times were 24.7 minutes
(SD = 6.2) and 32.2 minutes (SD = 4.5), respectively.
Completion times ranged from 16 to 34 minutes with \systemname{}
and from 26 to 40 minutes with Baseline.

The observed completion rate was identical across conditions,
while completion times were descriptively lower with \systemname{}
among completers. This comparison excludes the two incomplete
tasks and should be interpreted together with the completion
counts; it does not establish a statistically significant
difference in completion time.

\subsection{Evidence Quality}

Evidence quality was scored on a 0--2 scale.
Seven participants in the \systemname{} condition received
a score of 2, two received a score of 1, and one received
a score of 0. In the Baseline condition, six participants
received a score of 2, three received a score of 1, and
one received a score of 0. The median score was 2 in
both conditions.

The distributions were similar, with one additional
participant receiving the highest score in the \systemname{}
condition. This aggregate measure does not separately
establish the quality of skill localization, evidence-chain
reconstruction, or maintenance planning.

\subsection{Revisions and Final Decisions}

Participants in both conditions were able to edit a skill and run
the revised version. The main difference appeared in how the
results of those runs were used when deciding whether to retain
the revision.

In the \systemname{} condition, participants checked the result of the revision against other affected cases
as well as the case that motivated the change. When the selected
case improved, they also examined whether other cases improved
or regressed before making the final decision. The improvement
of the selected case was therefore not used by itself to justify
retaining the revision.

In the Baseline condition, participants retained
a revision after it improved the selected case or a small set of cases. In these instances, the revision addressed the immediate
failure, but a larger error pattern affecting more cases in the
same comparison remained unresolved. We also observed cases in which a favorable rerun was treated as
confirmation of the earlier explanation even though the available
evidence did not rule out other explanations.

These observations do not show that \systemname{} produced better
final skill versions. The difference was instead in what
participants treated the rerun as supporting. Participants using
\systemname{} checked the effect of a revision beyond the case
that motivated it before retaining the change, whereas several
Baseline participants retained revisions on the basis of a
narrower improvement.

\subsection{Usability and Subjective Experience}

\paragraph{Perceived usability.}

%Figure~\ref{fig:sus-results} presents the distribution of System Usability Scale (SUS) scores. \systemname{} received a mean SUS score of 70.00, slightly above the commonly used normative reference mean of 68 \cite{sauro2011practical}. This places its mean perceived usability slightly above that reference for the tasks evaluated in this study. The reference is used descriptively rather than as a pass--fail threshold or evidence of a statistically significant departure from the normative mean.

%Individual \systemname{} scores extended both below and above the reference value, indicating variation in participants' experiences. Baseline scores are included in the figure for context. SUS measures perceived usability and does not independently establish diagnostic accuracy or the quality of maintenance decisions.

On the System Usability Scale (SUS), \systemname{} received a mean score of 70.0, compared with 47.25 for the Baseline. Relative to the commonly used reference mean of 68~\cite{sauro2011practical}, the mean score for \systemname{} was slightly above the reference. Individual scores ranged from 57.5 to 77.5 for \systemname{} and from 27.5 to 80.0 for the Baseline, with greater variability in the Baseline condition. Overall, participants reported favorable perceived usability for \systemname{} during the skill-refinement task.

% \begin{figure}[t]
%     \centering
%     \includegraphics[width=\columnwidth]
%         {Figures/sus.png}
%     \Description{Violin plots of SUS scores for \systemname{} and Baseline, with individual participant points and condition means.}
%     \caption{Distribution of SUS scores by condition
%     (10 participants per condition).
%     Points represent individual participants, violin shapes
%     show the estimated score distributions, and green
%     diamonds indicate condition means.
%     Higher scores indicate greater perceived usability.}
%     \label{fig:sus-results}
% \end{figure}

\paragraph{Perceived workload.}

\begin{figure*}[t]
    \centering
    \includegraphics[width=\textwidth]
        {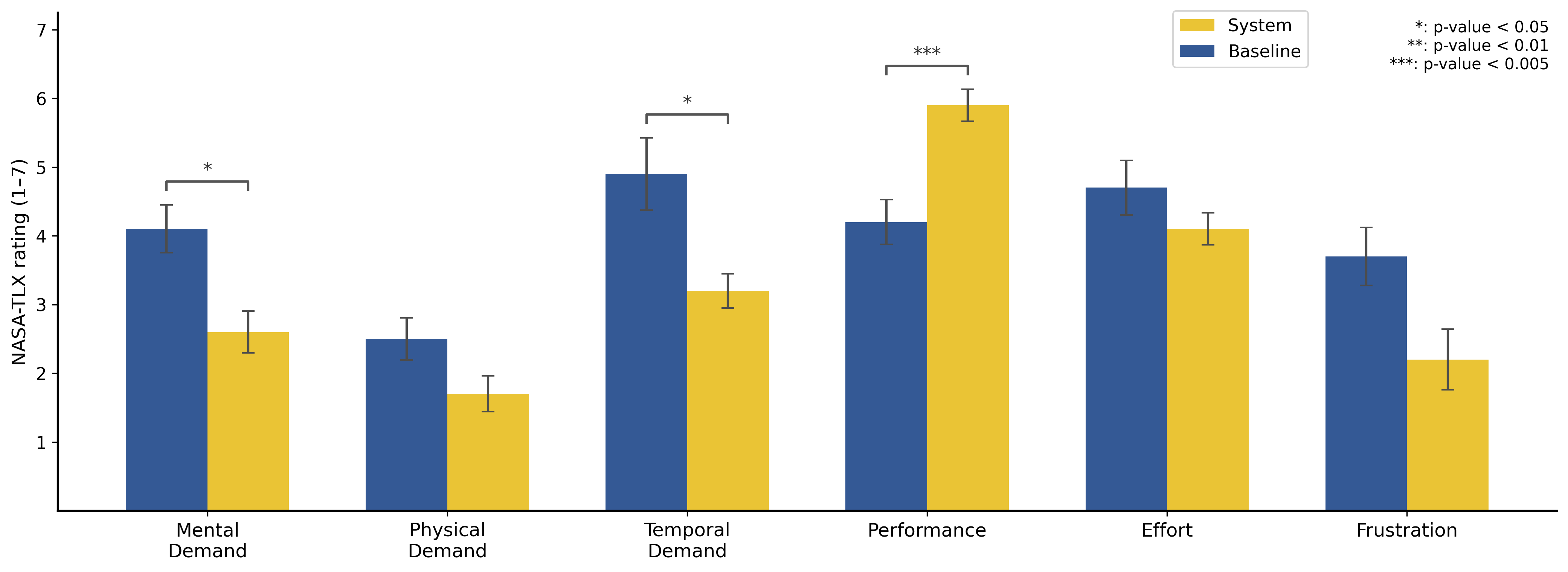}
    \Description{Grouped bar chart of six NASA-TLX dimensions for \systemname{} and Baseline, with significance brackets on mental demand, temporal demand, and performance.}
    \caption{Ratings for the six dimensions of the adapted seven-point NASA-TLX questionnaire by condition. Blue bars represent Baseline and yellow bars represent \systemname{}; error bars indicate SEM. Significance annotations are based on Holm-adjusted p-values across the six comparisons. Higher Performance ratings indicate better perceived performance; this item was direction-adjusted during data collection and was not reverse-coded during analysis.}
    \label{fig:nasa-tlx-results}
\end{figure*}

Figure~\ref{fig:nasa-tlx-results} presents the six NASA-TLX dimensions. After Holm correction, \systemname{} participants reported lower Mental Demand ($M = 2.60$ vs.\ $4.10$, $p_{\mathrm{Holm}} = .023$) and Temporal Demand ($M = 3.20$ vs.\ $4.90$, $p_{\mathrm{Holm}} = .048$), as well as higher direction-adjusted Performance ratings ($M = 5.90$ vs.\ $4.20$, $p_{\mathrm{Holm}} = .004$). Differences in Physical Demand, Effort, and Frustration were not statistically significant.

% Before submission, document the independent-samples test,
% effective sample size for each dimension, multiple-comparison
% correction, adjusted p-values, and the meaning of the error bars.
% Update the annotations if the statistical analysis changes.

\paragraph{Task confidence.}

\begin{figure*}[t]
    \centering
    \includegraphics[width=\textwidth]
        {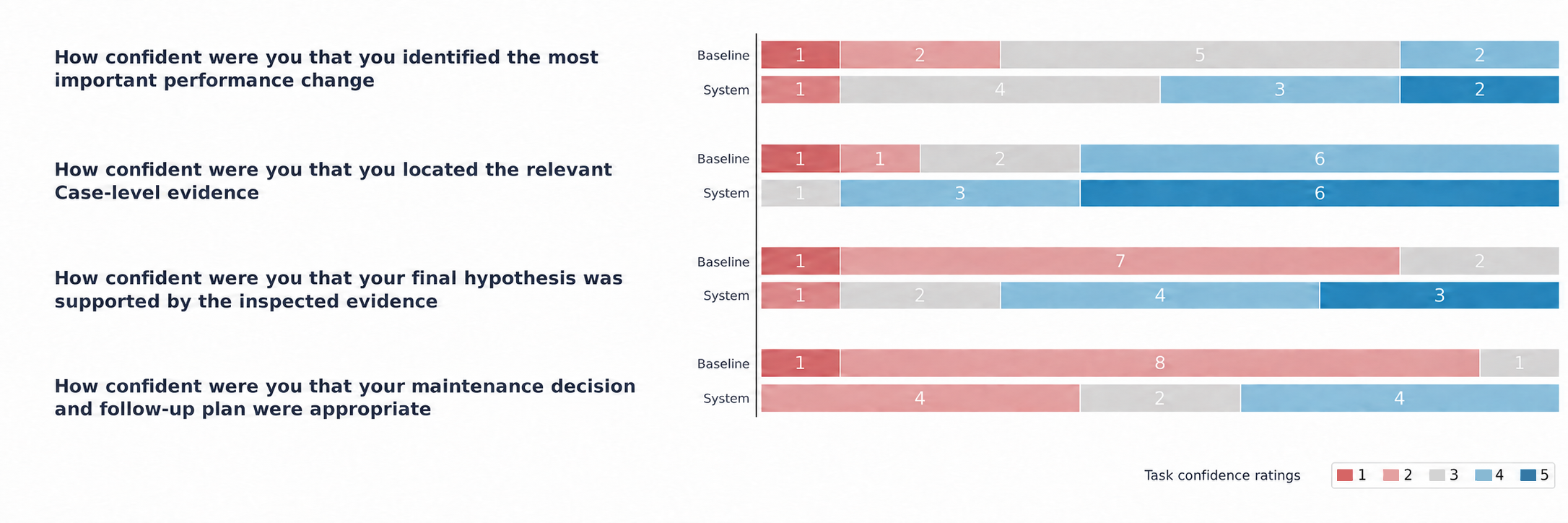}
    \Description{Stacked bars of 10 task-confidence ratings per condition for four items: performance-change identification, case-level evidence localization, evidence-supported hypothesis formation, and maintenance planning.}
    \caption{Task confidence ratings by condition.
    Each stacked bar contains 10 responses, and segment
    labels indicate participant counts.
    Colors represent ratings from 1 to 5, with higher
    values indicating greater confidence.
    The four items concern performance-change identification,
    case-level evidence localization, evidence-supported
    hypothesis formation, and maintenance planning.}
    \label{fig:task-confidence}
\end{figure*}

Figure~\ref{fig:task-confidence} shows the distributions
of participants' confidence ratings for four aspects
of the diagnostic and maintenance task.
We describe ratings of 4 or 5 as high-confidence responses.
For identifying the most important performance change,
5 of 10 \systemname{} participants reported high confidence,
compared with 2 of 10 Baseline participants.
For locating relevant case-level evidence, the corresponding
counts were 9 of 10 and 6 of 10.
For confidence that the final hypothesis was supported
by the inspected evidence, 7 of 10 \systemname{} participants
selected ratings of 4 or 5, whereas no Baseline participant
selected either rating.
For the appropriateness of the maintenance decision and
follow-up plan, 4 of 10 \systemname{} participants reported
high confidence, compared with none in the Baseline
condition.
These distributions show descriptively higher task
confidence with \systemname{}, particularly for forming
an evidence-supported hypothesis.
However, confidence in maintenance planning remained
mixed: 6 of 10 \systemname{} participants selected ratings
of 3 or below.
This suggests that deciding what to do after inspecting
the evidence remained challenging for some participants.
We therefore treat these ratings as a subjective complement
to the revision and run records above, rather than as evidence
that the final decisions were better.

% We need to discuss: whether and how our system deals with the overfitting problem.

\section{Discussion}
\label{sec:discussion}

\subsection{From Debugging Artifacts to Longitudinal Sensemaking}

Skill maintenance is not simply a matter of locating an error and editing a rule. Developers must connect an observed outcome change to the skill, invocation sequence, case, and execution evidence that make the change interpretable, then carry that interpretation into a later maintenance action. These materials are usually distributed across version histories, result tables, traces, and outputs, so the main difficulty is maintaining their relationships while the investigation moves between levels of detail.

\systemname{} treats this work as a longitudinal sensemaking process. A developer can begin with an outcome transition, narrow it to a skill or ordered sequence, return to the motivating case, inspect its paired Trace, Skill Diff, and Output, and then formulate a bounded What-if plan. The value of this chain is not that it produces a final explanation automatically. It lets developers progressively narrow the problem while retaining the scope and evidence that shaped the current interpretation. This comparison-local organization also distinguishes \systemname{} from tools that focus on a single execution or on generating a new capability: the unit of analysis remains connected across versions, cases, and maintenance rounds.

\subsection{Local Repair versus Family-Level Generalization}

Versioned Skills create a recurring tension between repairing a motivating case and preserving behavior across the broader skill family. A local fix can become overly specific to a case's wording, output format, tool path, or trigger keyword. Repeatedly appending such rules can also lengthen the skill, create conflicts, narrow its applicability boundary, or alter which skill is selected. A successful stochastic run may further make a local change appear more reliable than the available evidence supports. These are maintenance risks because a revision can encode the motivating case more specifically without improving behavior on other cases.

This tension also appeared in our evaluation. Several Baseline
participants retained a revision after it improved one or a few
selected cases, even though a larger error pattern affecting more
cases remained unresolved.
\systemname{} keeps this tension visible by retaining improvements and regressions in the same comparison context. A developer can carry the motivating case together with counterexample or held-out cases into a hypothesis and declare what should change, what should remain fixed, what should be measured, and what observation would weaken the proposal. Sample, Full, and Multiple Runs then provide progressively broader checks, while append-only hypotheses and decisions preserve why the change was made and how later evidence affected it. The system therefore turns generalization from an implicit assumption into an explicit maintenance question.

The distinction is deliberately simple. ``This case improved'' means that the selected case changed under the tested conditions. ``This skill generalizes better'' requires evidence from other cases, counterexamples, held-out scope, or repeated runs. A local improvement is evidence about the selected case under the tested conditions; broader generalization remains a question for cross-case and repeated-run examination. This framing also gives visual analytics a concrete role in skill maintenance: it can help developers respond to local failures without silently overfitting the maintained artifact to the case that first exposed the problem.

\subsection{Progressive Formalization in Human--LLM Maintenance}

The workflow uses different levels of structure at different points in an investigation. During exploration, transition summaries, skill and sequence patterns, and model-generated descriptions help developers decide what to inspect first. At this stage, concise abstraction is useful because the goal is to reduce the search space. During interpretation, the developer returns to paired case records, Traces, Skill Diffs, Outputs, and Sources. The candidate explanation must remain linked to those records, and the interface must make the evidence scope and provenance visible. During action, the developer confirms or edits a hypothesis and specifies a bounded What-if plan. The proposal becomes more structured because it is about to authorize a new run.

This progression separates the roles of the system and the person. The LLM can scan, group, summarize, and propose; the developer inspects, revises, authorizes, and decides. \systemname{} supports this division by keeping model suggestions editable and by requiring a human-confirmed classification before a refinement plan can be created. The result is a form of progressive formalization: an open-ended observation can become a candidate association, a reviewed interpretation, a hypothesis, and finally a bounded test without treating these states as equivalent. The evaluation also shows why these states should remain separate:
a favorable rerun can strengthen an explanation without, by itself,
showing that the revision should be retained for the broader skill.
This framing also follows established accounts of sensemaking as searching for and revising representations to make information easier to work with \cite{russell1993sensemaking}, and analytic-provenance work that captures and revisits the history of analysis \cite{nguyen2016sensepath,ragan2016survey}.

\subsection{Recoverable Maintenance as a Design Implication}

Longitudinal maintenance requires more than preserving the current recommendation. It requires preserving how the recommendation emerged: what changed, which cases motivated the change, which evidence supported it, what remained uncertain, what was modified in the skill, and what happened after a new run. In \systemname{}, comparison references, case and skill scopes, evidence links, hypotheses, revisions, and attached runs form a recoverable chain rather than a sequence of disposable screens.

This recoverability has three design consequences. First, context should be preserved while evidence narrows, so that Performance, Summary, Context, and Refinement remain stages of one investigation. Second, uncertainty should remain visible while commitment increases: an observation, a candidate association, a reviewed interpretation, and an authorized test should be distinguishable states. Third, decisions should remain available for later iterations, allowing a developer to challenge an earlier classification, revisit a case, or compare a new run without erasing the rationale for the previous one. For longitudinal maintenance, recoverability is therefore part of analytical quality: a useful system should preserve not only the current decision, but also the evidence and scope from which that decision emerged.
\subsection{Limitations and Future Work}
Our findings should be interpreted in light of four limitations. First, the between-subjects study included 20 participants, with 10 participants per condition, limiting the statistical power and precision of the estimated between-condition differences. We therefore interpret the results together with effect sizes and confidence intervals and avoid broad population-level claims. Second, the study used a fixed subset of FRAMES and a single web-search skill family. Although this setting exercised version comparison, evidence inspection, and rerun-based maintenance, it may not represent skills for coding, data analysis, multi-modal tasks, or other agent architectures. Third, the 40-minute laboratory task compressed a maintenance process that may unfold over days or weeks and therefore cannot establish whether the observed benefits persist during long-term use. Finally, participants selected different Base–Target intervals and maintenance targets. This supported self-directed investigation but introduced variation in task difficulty and reduced direct comparability across participants. Future work should evaluate \systemname{} with larger samples, additional skill families and benchmarks, and longitudinal deployments combining standardized and self-directed maintenance tasks.

\section{Conclusion}

Versioned agent Skills create a recurring maintenance problem: a revision can change outcomes, routing, and behavior across cases, while the evidence needed to understand that change is distributed across runs, Traces, skill Diffs, and Outputs. \systemname{} addresses this problem with a recoverable workflow that moves from performance change to case transition, skill and sequence evidence, original records, a human-authored hypothesis, and a bounded What-if test. Its central contribution is to preserve comparison and case scope as developers move between aggregate patterns, detailed evidence, and later maintenance rounds.

The workflow does not select a best skill automatically. It gives developers a way to connect local improvements to their evidence, examine them against other cases, and revisit them as later runs reveal how broadly a revision applies. In this sense, \systemname{} turns skill evolution into an inspectable refinement process in which local repair and family-level generalization remain separate questions, and each new decision can be challenged, refined, and tested in context.

\bibliographystyle{ACM-Reference-Format}
\bibliography{main}

\end{document}